\documentclass[journal]{IEEEtran}
\IEEEoverridecommandlockouts
\usepackage{amsmath,amssymb,amsfonts}
\usepackage{cite}

\allowdisplaybreaks[4]
\usepackage{algorithmic}
\usepackage{stfloats}
\usepackage{graphicx}
\usepackage{textcomp}
\usepackage{xcolor}
\usepackage{color}
\usepackage{pifont}
\usepackage{amsthm}

\usepackage{hyperref}
\usepackage{mathrsfs}
\usepackage{booktabs}
\usepackage{hyperref}
\usepackage{cases}
\usepackage{comment}
\usepackage{empheq} 
\usepackage{caption}
\usepackage{subcaption}
\usepackage{url}  
\usepackage{graphicx}  
\usepackage[ruled,vlined,linesnumbered]{algorithm2e}
\usepackage{setspace}
\usepackage{floatpag} 
\usepackage{float}
\usepackage{cite}
\usepackage{amsmath,amssymb,amsfonts}
\usepackage{algorithmic}
\usepackage{graphicx}
\usepackage{textcomp}
\usepackage{xcolor}
\renewcommand{\baselinestretch}{0.97}
\def\BibTeX{{\rm B\kern-.05em{\sc i\kern-.025em b}\kern-.08em
    T\kern-.1667em\lower.7ex\hbox{E}\kern-.125emX}}
\begin{document}

\title{Improving Physical-Layer Security in ISAC-UAV System: Beamforming and Trajectory Optimization\\
}

\author{Yue Xiu,~Wanting Lyu,~Phee~Lep Yeoh,~\IEEEmembership{senior Member,~IEEE}, Yi Ai\\~Yonghui Li,~\IEEEmembership{Fellow,~IEEE},
~Ning Wei,~\IEEEmembership{Member,~IEEE}\\
\thanks{Y.~Xiu is with College of Air Traffic Management, 
Civil Aviation Flight University of China, Deyang 618311, China (E-mail:  
xiuyue12345678@163.com). W.~Lyu and N.~Wei are with National Key Laboratory of Science and Technology on Communications, University of Electronic Science and Technology of China, Chengdu 611731, China (E-mail:  
lyuwanting@yeah.net, wn@uestc.edu.cn).
Phee Lep Yeoh is with School of Science, Technology and Engineering, University of the Sunshine Coast, QLD, Australia(e-mail: pyeoh@usc.edu.au).
Yi Ai is with College of Air Traffic Management, 
Civil Aviation Flight University of China, Deyang 618311, China, and School of Science, Technology and Engineering, Southwest Jiaotong University, Cheng Du, China (e-mail: aiyi@my.swjtu.edu.cn).
Yonghui Li is with the School of Electrical and Information Engineering, University of Sydney, Sydney, NSW 2006, Australia(e-mail: yonghui.li@sydney.edu.au).
}
\thanks{The corresponding author is Ning Wei.}}

\maketitle

\begin{abstract}
This paper investigates a novel unmanned aerial vehicle (UAV) secure communication system with integrated sensing and communications. We consider wireless security enhancement for a multiple-antenna UAV transmitting ISAC waveforms to communicate with multiple ground Internet-of-Thing devices and detect the surrounding environment. Specifically, we aim to maximize the average communication secrecy rate by optimizing the UAV trajectory and beamforming vectors. Given that the UAV trajectory optimization problem is non-convex due to multi-variable coupling develop an efficient algorithm based on the successive convex approximation (SCA) algorithm. Numerical results show that our proposed algorithm can ensure the accuracy of sensing targets and improve the communication secrecy rate. 
\end{abstract}


\section{Introduction}
In recent years, with the development of autonomous driving and artificial intelligence, demands for data traffic in wireless communications have increased exponentially, which poses a significant challenge to the development of wireless communication systems. 6-th generation (6G) wireless systems are required to support secure low-latency communications to a wide range of advanced applications, including the emerging Internet-of-Things (IoT) networks. To provide flexible coverage for a large number of ground IoT devices, a promising solution is to deploy unmanned aerial vehicle (UAV) base stations (BSs). Compared with traditional ground BSs, UAV BSs are not restricted by terrain and can be deployed case-by-case to improve communication quality. In addition, the transmission between UAVs and ground IoT devices is mainly based on line of sight (LoS) since UAVs typically operate at high altitudes. Despite these advantages, communication security remains challenging for UAV communications since other untrusted devices could directly eavesdrop on the UAV's LoS transmissions \cite{9954169,10622582,10284919}.

In advanced 6G IoT applications, BSs are expected to provide both communication and target sensing capabilities to obtain IoT location information. This has led to the emergence of a new wireless communication concept known as Integrated Sensing and Communication (ISAC) \cite{9606831}. The mutual benefits derived from the coordination of sensing and communication have obtained significant attention from both academia and industry \cite{9737357,5440129}.
\textcolor{blue}{However, in UAV and ISAC systems, the openness of the channels makes information susceptible to eavesdropping and attacks. Traditional security methods primarily rely on encryption technologies, but these often require additional computational resources and may face the risk of being compromised as computational capabilities advance. The emergence of Physical Layer Security (PLS) offers technological support for enhancing the security of wireless communications, and it has gained widespread attention in recent years. PLS focuses on leveraging the randomness of wireless channels and physical characteristics like multipath propagation to achieve secure transmission. Numerous studies have already explored the application of PLS in communication systems. Specifically, in\cite{8533374},
the authors developed a novel hierarchical physical layer security model to protect multiple messages simultaneously. This model maintains a layered security structure, where each message transmitted from the base station is assigned a security level, and each user in the network possesses a corresponding security clearance. In\cite{9672766}, the authors considered a cellular-connected UAV (C-UAV) serviced by a large-scale multiple-input multiple-output (MIMO) link to extend coverage while simultaneously enhancing physical layer security and authentication. A Rician channel was considered, and a novel linear precoder design was proposed for the transmission of both data and artificial noise.}
Moreover, several studies \cite{10168298,li2024uav,10529955,son2024secrecy,10054167} have explored Physical Layer Security (PLS) techniques in ISAC-UAV systems, where the flexibility of UAVs offers increased degrees of freedom (DoFs) for enhancing security and sensing performance. For instance, in \cite{10168298,li2024uav}, the authors proposed an semidefinite programming (SDP) relaxation to interfere with a ground eavesdropper (Eve) while improving target sensing accuracy and overall secrecy rates. Additionally, in \cite{10529955,son2024secrecy}, the authors employed the extended Kalman filter technique to predict user motion states, enabling the joint design of radar signals and receiver beamforming to enhance the secrecy rate. In \cite{10054167}, the authors introduced a novel adaptive ISAC for UAV systems, allowing UAVs to sense on-demand during communication processes. The sensing duration is flexibly configured based on application requirements, thus avoiding over-sensing and radio resource waste while improving resource utilization and system secrecy rate.

\textcolor{blue}{However, we noted that the existing references often consider Eves and sensing targets separately, whereas, in practice, Eves may disguise themselves as legitimate targets. Moreover, the above studies evaluated sensing performance based on the Cramér-Rao Bound (CRB). Although sensing accuracy, as measured by the CRB, is critical for system precision, the advantages of sensing rate may be more pronounced in UAV communication scenarios. This is particularly true in applications that require real-time performance, rapid response, adaptation to dynamic environments, reduced latency, and support for multitasking\cite{9800940}. However, the existing references do not study the ISAC-UAV system under sensing rate constraints.} To address these gaps, our paper examines a ISAC-UAV system, where the UAV communicates with ground IoT devices while sensing an untrusted target (UT), which is treated as a potential Eve. Then, we formulate an optimization problem to design the UAV beamforming vector and the trajectory. Due to the coupling of these optimization variables, the problem is non-convex. To tackle this challenge, we propose an efficient algorithm based on the successive convex approximation (SCA) method. Simulation results demonstrate that the proposed algorithm significantly improves the secrecy rate as the UAV transmit power increases. We also highlight the trade-off between meeting the secrecy and sensing rates in the optimized ISAC-UAV trajectory, providing valuable insights for future ISAC-UAV systems.

\section{System Model and Problem Formulation}\label{II}
\begin{figure}[htbp]
  \centering
  \includegraphics[scale=0.4]{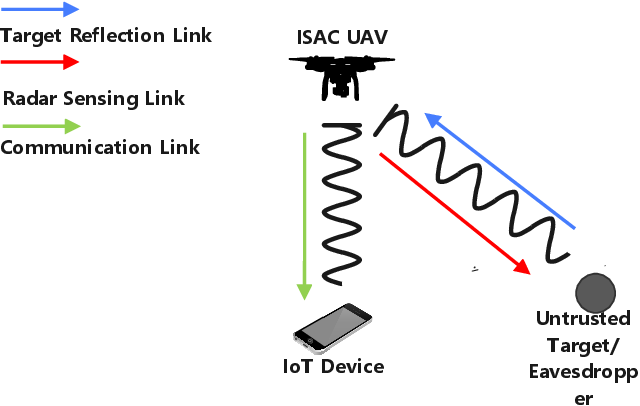}
  \caption{System model of UAV with ISAC sending communications to the IoT device and sensing the location of the UT. The UT is also considered as an Eve attempting to access the communications sent to the IoT device.}
\label{FIG1}
\end{figure}
We consider a secure ISAC-UAV system, as shown in Fig.\ref{FIG1}, {where the UAV is communicating with the IoT device and sensing the location of the untrusted target (UT) using the same ISAC waveform}. In this system, the ISAC-UAV is equipped with a uniform linear array (ULA) {which} has $N_{t}$ transmitting antennas and $N_{r}$ sensing receiving antennas. The locations of {the} IoT device and {UT} are expressed  {according to a 3D coordinate system as} {$\mathbf{p}_{d}[s]=[x_{d}[s], y_{d}[s],0]^{T}$ and $\mathbf{p}_{t}[s]=[x_{t}[s], y_{t}[s],0]^{T}$}, respectively. {To perform the UAV trajectory optimization,} we {consider the UAV needs to complete the ISAC functions within a given time duration of $T$, and $T$ is divided into} $S$ slots, where $T=S\Delta_{s}$ and $\Delta_{s}$ represent the slot length, $s\in\{1,\ldots,S\}$ is the $s$-th time slot. The position of the UAV {is} expressed as $\mathbf{p}_{u}[s]=[x_{u}[s], y_{u}[s], z_{u}]^{T}$, where the UAV {is} flying horizontally, so the height of the UAV remains fixed.
\subsection{Communication Signal Model}
{For the communication link, the ISAC signal} $x[s]$ sent by the UAV reaches the IoT device and {UT} {in each} $s$-th time slot. The signals received at {the} IoT device, and UT, are respectively expressed as
\begin{small}
\begin{align}
&y_{ud}[s]=L_{ud}[s]\mathbf{h}_{ud}^{H}[s]\mathbf{w}[s]x[s]+n_{d}[s],\label{pro1}\\
&y_{ut}[s]=L_{ut}[s]\mathbf{h}_{ut}^{H}[s]\mathbf{w}[s]x[s]+n_{t}[s],\label{pro3}
\end{align}    
\end{small}%
where {$n_{d}[s]\sim\mathcal{CN}(0,\sigma^{2}_{d})$ and $n_{t}[s]\sim\mathcal{CN}(0,\sigma^{2}_{t})$}, $L_{ud}[s]=\sqrt{\rho(z_{u}^{2}+\|\bar{\mathbf{p}}_{u}[s]-\mathbf{p}_{d}[s]\|)^{-\frac{\kappa}{2}}}$, $L_{ut}[s]=\sqrt{\rho(z_{u}^{2}+\|\bar{\mathbf{p}}_{u}[s]-\mathbf{p}_{t}[s]\|)^{-\frac{\kappa}{2}}}$ denotes the distance-dependent path loss model of {the links from the} UAV-to-IoT device {and the UAV-to-UT}, respectively\cite{10168298}, where 
$\rho$ is the path loss at the reference distance $1$m. $\kappa$ is the path loss exponents for the UAV-to-IoT device link and the {UAV-to-UT} link, respectively, and $\bar{\mathbf{p}}_{u}[s]=[x_{u}[s],y_{u}[s]]^{T}$. According to \cite{li2024uav}, we assume all channel models for communication are Rician channel models. Hence, the small-scale fading channel models are $
\mathbf{h}_{ud}[s]=\sqrt{\frac{\beta_{ud}}{1+\beta_{ud}}}\mathbf{h}_{ud}^{LoS}[s]+\sqrt{\frac{1}{1+\beta_{ur}}}\mathbf{h}_{ud}^{NLoS}[s]$, $
\mathbf{h}_{ut}[s]=\sqrt{\frac{\beta_{ut}}{1+\beta_{ut}}}\mathbf{h}_{ut}^{LoS}[s]$ $+\sqrt{\frac{1}{1+\beta_{ut}}}\mathbf{h}_{ut}^{NLoS}[s]$, where $\beta_{ud}$, $\beta_{ut}$ are the Rician factor of the UAV-to-IoT device  and UAV-to-UT, respectively. $\mathbf{h}_{ud}^{LoS}[s]$ and $\mathbf{h}_{ut}^{LoS}[s]$ denote the channel LoS component. $\mathbf{h}_{ud}^{NLoS}[s]$ and {$\mathbf{h}_{ut}^{NLoS}[s]$ denote the channel NLoS component\cite{son2024secrecy}.

\subsection{Target Sensing Model}
In addition to transmitting the information from UAV to the IoT device, $x[s]$ is also used for target detection. As shown in Fig.~\ref{FIG1}, signal $x[s]$ goes through the {UAV-to-UT-to-UAV} link. Finally, the reflected echo signal is collected at the UAV. Moreover, according to \cite{9133130}, since the channel state information (CSI) of the IoT device is known, the reflected signal of the IoT device can be removed from the echo signal. Thus, the echo signal is expressed as follows
\begin{align}
r_{ut}[s]=\mathbf{u}^{H}L_{ut}^{2}[s]\mathbf{h}_{ut}\mathbf{h}_{ut}^{H}\mathbf{w}[s]x[s]+n_{ut}[s],\label{pro4}
\end{align}
where {$\mathbf{u}\in\mathbb{C}^{N_{r}\times1}$} is the receive beamformer at the UAV and $\|\mathbf{u}\|_{2}=1$. $n_{ut}[s]\sim\mathcal{CN}(0,\sigma^{2}_{ut})$ is the Additive white Gaussian noise (AWGN) at the UAV.

\subsection{Performance Metrics of Communication, Security and Sensing}
{To analyze the performance of the communication link, we derive} the signal-to-noise ratio (SNR) of the IoT device based on (\ref{pro1}) as
\begin{small}
\begin{align}
\mathrm{SNR}_{ud}[s]=|L_{ud}[s]\mathbf{h}_{ud}^{H}[s]\mathbf{w}[s]|^{2}/{\sigma^{2}}.\label{pro6}
\end{align}
\end{small}%
{Based on (\ref{pro4})}, the SNR of the UT is {derived as}
\begin{small}
\begin{align}
\mathrm{SNR}_{ut}[s]=|L_{ut}[s]\mathbf{h}_{ut}[s]^{H}\mathbf{w}[s]|^{2}/{\sigma^{2}}.\label{pro9}
\end{align}
\end{small}%
Next, we obtain the communication secrecy rate of the ISAC-UAV system which is given by
\begin{small}
\begin{align}
&\bar{R}_{sec}[s]=[\log_{2}(1+\mathrm{SNR}_{ud}[s])-\log_{2}(1+\mathrm{SNR}_{ut}[s])]^{+},\label{pro10}
\end{align}
\end{small}%
Then, the average secrecy rate is written as
\begin{small}
\begin{align}
R_{sum}[s]=\frac{1}{S}\sum_{s=1}^{S}\bar{R}_{sec}[s].\label{pro20}
\end{align}
\end{small}%
where $[x]^{+}=\max(0, x)$. For the sensing performance, we derive the SNR of the UT echo signal based on (\ref{pro3}) as
\begin{small}
\begin{align}
\widetilde{\mathrm{SNR}}_{ut}[s]=\frac{|\mathbf{u}[s]^{H}L_{urt}^{2}[s]\mathbf{h}_{ut}\mathbf{h}_{ut}^{H}\mathbf{w}[s]|^{2}}{\mathbf{u}[s]^{H}\mathbf{u}[s]\sigma^{2}}.\label{pro_9}
\end{align}
\end{small}%
Therefore, the optimization problem can be expressed as
\begin{small}
\begin{subequations}
\begin{align}
\max_{\bar{\mathbf{p}}_{u}[s],\mathbf{w}[s],\mathbf{u}[s]}&~R_{sum}[s],\label{pro21a}\\
\mbox{s.t.}~
&\|\bar{\mathbf{p}}_{u}[s+1]-\bar{\mathbf{p}}_{u}[s]\|^{2}\leq D^{2}, s=1,\ldots,S-1,&\label{pro21c}\\
&\|\bar{\mathbf{p}}_{u}[S]-\bar{\mathbf{p}}_{u}[F]\|^{2}\leq D^{2}, \bar{\mathbf{p}}_{u}[1]=\bar{\mathbf{p}}_{u}[0],&\label{pro21d}\\
&\widetilde{\mathrm{SNR}}_{ut}[s]\geq \gamma,&\label{pro21g}\\
&\|\mathbf{w}[s]\|\leq P,&\label{pro21h}\\
&\|\mathbf{u}[s]\|_{2}= 1.&\label{pro21i}
\end{align}\label{pro21}%
\end{subequations}
\end{small}%
(\ref{pro21c}) and (\ref{pro21d}) are the mobility constraints of the UAV, where $\bar{\mathbf{p}}_{u}[0]$ and $\bar{\mathbf{p}}_{u}[F]$ are the initial and final positions of the UAV. $D=v_{max}\Delta_{s}$ denotes the maximum distance of the UAV in $\Delta_{s}$, where $v_{max}$ is the maximum speed of the UAV. Constraint (\ref{pro21g}) is to guarantee the sensing performance with $\gamma$ as the minimum required SNR of the echo signal at the UAV. (\ref{pro21h}) is the transmit power constraint, and $P$ is the maximum transmit power. (\ref{pro21i}) is the received beamforming constraint. Since both the objective function in (\ref{pro21}) and the constraints in (\ref{pro21}) are non-convex, the problem in (\ref{pro21}) is non-convex. To solve the above non-convex problem, we propose an SCA-based algorithm to solve the above problem by optimizing the transmit beamforming vector, UAV trajectory, and receive beamforming vector.

\section{ISAC-UAV Secrecy Rate Optimization}\label{III}
\subsection{{ISAC-UAV} Transmit Beamforming Optimization}
{We first optimise the ISAC transmit beamforming for a fixed UAV trajectory and receive beamforming vector. As such, given $\mathbf{u}[s]$, $\bar{\mathbf{p}}_{u}[s]$, where $\bar{\mathbf{p}}_{u}[s]$ is the UAV's trajectory variable. The optimization in (\ref{pro21}) simplifies to}
\begin{small}
\begin{subequations}
\begin{align}
\max_{\mathbf{w}[s]}&~R_{sum}[s],\label{pro201_a}\\
\mbox{s.t.}~
&(\ref{pro21g}),(\ref{pro21h}).&\label{pro201b}
\end{align}\label{pro201}%
\end{subequations}
\end{small}%
\textcolor{blue}{The core idea of the SCA algorithm is to transform a complex non-convex optimization problem into a series of more manageable convex problems through successive convex approximations. The algorithm begins each iteration by creating a local convex approximation of the non-convex objective function and constraints at the current point. It then solves this approximated convex problem to update the solution. Therefore,}
to perform the SCA optimization, we introduce {a} slack variable $ \bar{\gamma}$, and we have $\log_{2}(1+\mathrm{SNR}_{ud}[s])\geq \bar{\gamma}$. Problem (\ref{pro201}) is rewritten as
\begin{small}
\begin{subequations}
\begin{align}
\max_{\mathbf{w}[s],\bar{\gamma}}&~\bar{\gamma}-\frac{1}{S}\sum_{s=1}^{S}\log_{2}(1+\mathrm{SNR}_{ut}[s]),\label{pro201a}\\
\mbox{s.t.}~
&\log_{2}(1+\mathrm{SNR}_{ud}[s])\geq \bar{\gamma},&\label{pro203b}\\
&(\ref{pro21g}),(\ref{pro21h}).&\label{pro203c}
\end{align}\label{pro203}%
\end{subequations}
\end{small}%
{We also use the approximation of the} first order Taylor series of (\ref{pro203b})
\begin{small}
\begin{align}
&2\mathrm{Re}((\mathbf{h}_{ud}^{H}[s]\mathbf{w}_{0}[s])^{H}\mathbf{h}_{ud}^{H}[s]\mathbf{w}[s])-|\mathbf{h}_{ud}^{H}[s]\mathbf{w}_{0}[s]|^{2}\geq \nonumber\\
&(2^{\bar{\gamma}}-1)\sigma^{2}/L_{ud}^{2},\label{pro2_1}
\end{align}
\end{small}%
{and the} {first order} Taylor series of (\ref{pro21g}) is expressed as
\begin{small}
\begin{align}
&2\mathrm{Re}((\mathbf{u}[s]^{H}[s]\mathbf{h}_{ut}\mathbf{h}_{ut}^{H}\mathbf{w}_{0}[s])^{H}\mathbf{u}[s]^{H}\mathbf{h}_{ut}\mathbf{h}_{ut}^{H}\mathbf{w}[s])\nonumber\\
&-|\mathbf{u}[s]^{H}\mathbf{h}_{ut}\mathbf{h}_{ut}^{H}\mathbf{w}_{0}[s]|^{2}\geq 2^{\gamma}\sigma^{2}/L_{urt}^{2}[s].\label{pro2_2}
\end{align}
\end{small}%
{Finally, the optimization in} (\ref{pro201}) is rewritten as
\begin{small}
\begin{subequations}
\begin{align}
\max_{\mathbf{w}[s]}&~\bar{\gamma}-\frac{1}{S}\sum_{s=1}^{S}\log_{2}(1+\mathrm{SNR}_{ut}[s]),\label{pro202a}\\
\mbox{s.t.}~
&(\ref{pro2_1}),(\ref{pro2_2}),(\ref{pro21h}),&\label{pro202b}
\end{align}\label{pro202}%
\end{subequations}
\end{small}%
{which is} convex {and can be solved using} CVX.
\subsection{{ISAC-UAV} Trajectory Optimization}
{Next, we need to optimise the ISAC-UAV trajectory for given transmit and receive beamforming vectors.} Since $\mathbf{h}^{LoS}_{ud}[s]$ is complex and non-linear to UAV trajectory, which makes the UAV's trajectory optimization problem difficult, we {proceed by applying the following approximations}. {We use} the UAV trajectory of iteration $(j-1)$-th to approximate $\mathbf{h}^{LoS}_{ud}[s]$ in {the} $j$-th {iteration to approximate the SNR terms in (\ref{pro6}) and (\ref{pro9}) as}
\begin{small}
\begin{align}
&|L_{ud}[s]\mathbf{h}_{ud}^{H}[s]\mathbf{w}[s]|^{2}\approx\left[\sqrt{(d_{ud}[s])^{-\kappa}}\right]\left[\begin{matrix}
\mathbf{h}_{ud}^{H}[s]\mathbf{w}[s]\\
  \end{matrix}
  \right]\nonumber\\
&\times\left[\begin{matrix}
\mathbf{h}_{ud}^{H}[s]\mathbf{w}[s]\\
  \end{matrix}
  \right]^{H}\left[\sqrt{(d_{ud}[s])^{-\kappa}}\right]^{H}.\label{pro56}
\end{align}
\end{small}%
and
\begin{small}
\begin{align}
&|L_{ut}[s]\mathbf{h}_{ut}^{H}[s]\mathbf{w}[s]|^{2}\approx\left[\sqrt{(d_{ut}[s])^{-\kappa}}\right]\left[\begin{matrix}
\mathbf{w}^{H}[s]\mathbf{h}_{ut}[s]\\
  \end{matrix}
  \right]&\nonumber\\
&\times\left[\begin{matrix}
\mathbf{w}^{H}[s]\mathbf{h}_{ut}[s]\\
  \end{matrix}
  \right]^{H}\left[\sqrt{(d_{ut}[s])^{-\kappa}}\right]^{H},\label{pro58}
\end{align}
\end{small}%
where $d_{ud}=\sqrt{(x_{u}[s]-x_{d}[s])^{2}+(y_{u}[s]-y_{d}[s])^{2}+z_{u}^{2}}$ and $d_{ut}=\sqrt{(x_{u}[s]-x_{t}[s])^{2}+(y_{u}[s]-y_{t}[s])^{2}+z_{u}^{2}}$. To simplify this expression, we let $\boldsymbol{\Psi}_{1}=\left[\begin{matrix}
\mathbf{h}_{ud}^{H}[s]\mathbf{w}[s]\\
  \end{matrix}
  \right]$, $\boldsymbol{\Psi}_{2}=\left[\begin{matrix}
\mathbf{h}_{ut}^{H}[s]\mathbf{w}[s]\\
  \end{matrix}
  \right]$ {and the SNRs in the communication secrecy rate in (\ref{pro10}) can be approximated as}
\begin{small}
\begin{align}
&R_{ud}[s]\approx\log_{2}\left(1+[\sqrt{(d_{ud}[s])^{-\kappa}}]\boldsymbol{\Psi}_{1}\boldsymbol{\Psi}_{1}^{H}[\sqrt{(d_{ud}[s])^{-\kappa}}]^{H}\right),\nonumber\\
&R_{ut}[s]\approx\log_{2}\left(1+[\sqrt{(d_{ut}[s])^{-\kappa}}]\boldsymbol{\Psi}_{2}\boldsymbol{\Psi}_{2}^{H}[\sqrt{(d_{ut}[s])^{-\kappa}}]^{H}\right).\label{pro59}
\end{align} 
\end{small}%
Given $\mathbf{u}[s]$ and $\mathbf{w}[s]$, the problem in (\ref{pro21}) is rewritten as
\begin{small}
\begin{subequations}
\begin{align}
\max_{\bar{\mathbf{p}}_{u}[s]}&~\frac{1}{S}\sum_{s=1}^{S}\nonumber\\
& \left(\log_{2}\left(1+[\sqrt{(d_{ud}[s])^{-\kappa}}]\boldsymbol{\Psi}_{1}\boldsymbol{\Psi}_{1}^{H}\right.[\sqrt{(d_{ud}[s])^{-\kappa}}]^{H}\right)\nonumber\\
  &{\scriptstyle\left.-\log_{2}\left(1+[\sqrt{(d_{ut}[s])^{-\kappa}}]\boldsymbol{\Psi}_{2}\boldsymbol{\Psi}_{2}^{H}[\sqrt{(d_{ut}[s])^{-\kappa}}]^{H}\right)\right),}\label{pro60a}\\
\mbox{s.t.}~
&(\ref{pro21c}),(\ref{pro21d}),(\ref{pro21g}).&\label{pro60b}
\end{align}\label{pro60}%
\end{subequations}    
\end{small}%
We introduce slack variables $\boldsymbol{\zeta}$ problem (\ref{pro60}) is rewritten as
\begin{small}
\begin{subequations}
\begin{align}
\max_{\bar{\mathbf{p}}_{u}[s],\boldsymbol{\zeta}}&~\frac{1}{S}(\log_{2}(1+\zeta_{1}[s])-\log_{2}(1+\zeta_{2}[s])),\label{pro61a}\\
\mbox{s.t.}~
&[\sqrt{(d_{ud}[s])^{-\kappa}}]\boldsymbol{\Psi}_{1}\boldsymbol{\Psi}_{1}^{H}[\sqrt{(d_{ud}[s])^{-\kappa}}]^{H}\geq \zeta_{1}[s],&\label{pro61b}\\
&[\sqrt{(d_{ut}[s])^{-\kappa}}]\boldsymbol{\Psi}_{2}\boldsymbol{\Psi}_{2}^{H}[\sqrt{(d_{ut}[s])^{-\kappa}}]^{H}\leq\zeta_{2}[s],&\label{pro61d}\\
&\sqrt{(d_{ud}[s])^{-\kappa}}\geq\zeta_{3}[s], \sqrt{(d_{ut}[s])^{-\kappa}}\leq\zeta_{4}[s],&\label{pro61h}\\
&\sqrt{(d_{ut}[s])^{-\alpha}}\geq\gamma_{t}^{1/4},&\label{pro61k}
\end{align}\label{pro61}%
\end{subequations}
\end{small}%
where $\boldsymbol{\zeta}=\{\zeta_{1}[s],\zeta_{2}[s],\zeta_{3}[s],\zeta_{4}[s]\}$, $\gamma_{t}=\frac{\gamma\mathbf{u}^{H}\mathbf{u}\sigma^{2}}{d_{ut}^{-2\alpha}\rho^{2}|\mathbf{u}^{H}[s]\mathbf{h}_{ut}[s]^{H}\mathbf{h}_{ut}[s]\mathbf{w}[s]|^{2}}$.
First, we {approximate } the objective function in (\ref{pro61a}) {by deriving the} first order Taylor expansion of $\log_{2}(1+\zeta_{2}[s])$ {which is} denoted as
\begin{small}
\begin{align}
&\log_{2}(1+\zeta_{2}[s])\leq\log_{2}(1+\zeta_{2,0}[s])+\frac{1}{\ln2(1+\zeta_{2,0}[s])}\nonumber\\
&(\zeta_{2}[s]-\zeta_{2,0}[s]).\label{for7}
\end{align}
\end{small}%
Therefore, the lower bound of (\ref{pro61a}) is expressed as
\begin{small}
\begin{align}
&\frac{1}{S}(\log_{2}(1+\zeta_{1}[s])-\log_{2}(1+\zeta_{2}[s])\leq\log_{2}(1+\zeta_{2,0}[s])\nonumber\\
&+\frac{1}{\ln2(1+\zeta_{2,0}[s])}(\zeta_{2}[s]-\zeta_{2,0}[s])).\label{for8}
\end{align}
\end{small}%
(\ref{for8}) is convex, and (\ref{for8}) is used to replace (\ref{pro61a}).
{Next,} we address the non-convexity of {the} constraints {by deriving the} first order Taylor expansions of (\ref{pro61b}) and (\ref{pro61d}) {which} are expressed as
\begin{small}
\begin{align}
&-[\zeta_{1,0}[s]]\boldsymbol{\Psi}_{1}\boldsymbol{\Psi}_{1}^{H} [\zeta_{1,0}[s]]^{H}+2\mathrm{Re}\left([\zeta_{1,0}[s]]\boldsymbol{\Psi}_{1}\boldsymbol{\Psi}_{1}^{H}[\zeta_{1}[s]\right)\nonumber\\
&\geq \zeta_{1}[s],\label{for5}\\
&-[\zeta_{2,0}[s]]\boldsymbol{\Psi}_{2}\boldsymbol{\Psi}_{2}^{H}[\zeta_{2,0}[s]]^{H}+2\mathrm{Re}\left([\zeta_{2,0}[s]]\boldsymbol{\Psi}_{2}\boldsymbol{\Psi}_{2}^{H}[\zeta_{2}[s]\right)\nonumber\\
&\leq\zeta_{2}[s].\label{for6}
\end{align}
\end{small}%
Then, $\sqrt{(d_{ud}[s])^{-\kappa}}\geq\zeta_{3}[s]$, $\sqrt{(d_{ut}[s])^{-\alpha}}\leq\gamma_{t}^{1/4}$ and $\sqrt{(d_{ut}[s])^{-\kappa}}\leq\zeta_{4}[s]$ are written as
\begin{small}
\begin{align}
&((x_{u}[s]-x_{d}[s])^{2}+(y_{u}[s]-y_{d}[s])^{2}+z_{u}^{2})^{-\kappa/4}\geq\zeta_{3}[s],\label{for1}\\
&((x_{u}[s]-x_{t}[s])^{2}+(y_{u}[s]-y_{t}[s])^{2}+z_{u}^{2}[s])^{-\kappa/4}\leq\zeta_{4}[s],\label{for2}\\
&x_{u}^{2}[s]+x_{d}^{2}[s]-2x_{u}[s]x_{d}[s]+y_{u}^{2}[s]+y_{d}^{2}[s]-2y_{u}[s]y_{d}[s]\nonumber\\
&+z_{u}^{2}[s]-\gamma_{t}^{\frac{1}{16\alpha}}\leq 0,\label{for_2}
\end{align}
\end{small}%
{which} is convex. The expressions in (\ref{for1}) and (\ref{for2}) {can be further approximated} as
\begin{small}
\begin{align}
&x_{u}^{2}[s]+x_{d}^{2}[s]-2x_{u}[s]x_{d}[s]+y_{u}^{2}[s]+y_{d}^{2}[s]-2y_{u}[s]y_{d}[s]\nonumber\\
&+z_{u}^{2}[s]-\zeta_{3}[s]^{-\frac{4}{\kappa}}\leq 0,\label{for3}\\
&\zeta_{4}[s]^{-\frac{4}{\kappa}}-x_{u}^{2}[s]-x_{t}^{2}[s]+2x_{u}[s]x_{t}[s]-y_{u}^{2}[s]-y_{t}^{2}[s]\nonumber\\
&+2y_{u}[s]y_{t}[s]-z_{u}^{2}[s]\geq 0.\label{for4}
\end{align}
\end{small}%
Since (\ref{for3}) is a non-convex constraint, we use the first order Taylor expansion of $\zeta_{3}[s]^{-\frac{4}{\kappa}}$ to approximate (\ref{for3}) {which is given by}
\begin{small}
\begin{align}
\zeta_{3}[s]^{-\frac{4}{\kappa}}\geq \zeta_{3,0}[s]^{-\frac{4}{\kappa}}-\frac{4}{\kappa}\zeta_{3,0}[s]^{-\frac{4}{\kappa}-1}(\zeta_{3}[s]-\zeta_{3,0}[s]).\label{for9}
\end{align}
\end{small}%
Based on (\ref{for9}), constraint (\ref{for3}) is rewritten as
\begin{small}
\begin{align}
&x_{u}^{2}[s]+x_{d}^{2}[s]-2x_{u}[s]x_{d}[s]+y_{u}^{2}[s]+y_{d}^{2}[s]-2y_{u}[s]y_{d}[s]\nonumber\\
&+z_{u}^{2}[s]-\zeta_{3,0}[s]^{-\frac{4}{\kappa}}+\frac{4}{\kappa}\zeta_{3,0}[s]^{-\frac{4}{\kappa}-1}(\zeta_{3}[s]-\zeta_{3,0}[s])\leq 0,\label{for10}
\end{align}
\end{small}%
{which} is convex. Similarly, the first order Taylor expansions of $-x_{u}^{2}$ and $-y_{u}^{2}$ are given by
\begin{small}
\begin{align}
&-x_{u}[s]^{2}\leq x_{u,0}^{2}[s]-2x_{u,0}[s]x_{u}[s], \nonumber\\
&-y_{u}^{2}[s]\leq y_{u,0}^{2}[s]-2y_{u,0}[s]y_{u}[s].
\end{align}
\end{small}%
Constraint (\ref{for4}) is rewritten as
\begin{small}
\begin{align}
&\zeta_{4}[s]^{-\frac{4}{\kappa}}+x_{u,0}^{2}[s]-2x_{u,0}[s]x_{u}[s]-x_{d}^{2}[s]+2x_{u}[s]x_{d}[s]y_{u,0}^{2}[s]\nonumber\\
&-2y_{u,0}[s]y_{u}[s]-y_{d}^{2}[s]+2y_{u}[s]y_{d}[s]-z_{u}^{2}[s]\geq 0,\label{for12}
\end{align}
\end{small}%
{which} is convex. 
{Finally}, the {ISAC-UAV trajectory optimization} in (\ref{pro61}) {can be} approximated as
\begin{small}
\begin{subequations}
\begin{align}
\max_{\bar{\mathbf{p}}_{u}[s],\boldsymbol{\zeta}}&~(\ref{for8}),\label{pro63a}\\
\mbox{s.t.}~
&,(\ref{for5}),(\ref{for6}),(\ref{for_2}),(\ref{for10}),(\ref{for12}),&\label{pro63b}
\end{align}\label{pro63}%
\end{subequations}
\end{small}%
{which} is convex.

\subsection{{ISAC-UAV} for Receiving Beamforming {Optimization}}
{Finally, we consider optimizing the ISAC receive beamforming for fixed UAV trajectory $\bar{\mathbf{p}}_{u}[s]$ and transmit beamforming vector $\mathbf{w}[s]$. We have}
\begin{small}
\begin{subequations}
\begin{align}
\max_{\mathbf{u}[s]}&~\frac{1}{S}\sum_{s=1}^{S}R_{sum}[s],\label{pro64a}\\
\mbox{s.t.}~
&(\ref{pro21g}),(\ref{pro21i}),&\label{pro64b}
\end{align}\label{pro64}%
\end{subequations}
\end{small}%
Based on \cite{7547360}, receive beamforming $\mathbf{u}[s]$ is not involved in $\mathrm{SNR}_{ud}[s]$, and the optimization objective is only relevant to $\mathrm{SNR}_{ut}[s]$. Thus, problem (\ref{pro64}) is rewritten as
\begin{small}
\begin{subequations}
\begin{align}
\min_{\mathbf{u}[s]}&~\frac{1}{S}\sum_{s=1}^{S}\mathbf{u}^{H}[s]\boldsymbol{\Omega}[s]\mathbf{u}[s],\label{pro65a}\\
\mbox{s.t.}~
&\mathbf{u}^{H}[s]\boldsymbol{\Omega}\mathbf{u}[s]\geq\gamma\sigma^{2},(\ref{pro21h}),&\label{pro65b}
\end{align}\label{pro65}%
\end{subequations}   
\end{small}%
where $\boldsymbol{\Omega}=L_{urt}^{4}[s]\mathbf{h}_{ut}[s]^{H}\mathbf{h}_{ut}[s]\mathbf{w}[s]\mathbf{w}[s]^{H}\mathbf{h}_{ut}[s]^{H}\mathbf{h}_{ut}[s]$.
According to \cite{7547360}, \textcolor{blue}{The core idea of the MM (Majorization-Minimization or Minorization-Maximization) algorithm is to iteratively approximate the solution to a complex original problem by constructing a simplified surrogate function. In each iteration, the algorithm selects a surrogate function that either tangentially touches or envelops the original objective function at the current iteration point but is more straightforward to optimize. A new iteration point is obtained by maximizing or minimizing this surrogate function, which provides a better solution or approximation to the original objective function. } we use the MM algorithm to resolve problem (\ref{pro65}) and problem (\ref{pro65}) is rewritten as
\begin{small}
\begin{subequations}
\begin{align}
\min_{\mathbf{u}[s]}&~\frac{1}{S}\sum_{s=1}^{S}\mathbf{u}^{H}[s]\lambda_{max}(\boldsymbol{\Omega})\mathbf{u}[s]+2\mathrm{Re}(\mathbf{u}^{H}[s](\boldsymbol{\Omega}-\lambda_{max}(\boldsymbol{\Omega}))\mathbf{I})\nonumber\\
&\mathbf{u}_{0}^{H}[s],\label{pro66a}\\
\mbox{s.t.}~
&\mathbf{u}^{H}[s]\lambda_{max}(\boldsymbol{\Omega})\mathbf{u}[s]+2\mathrm{Re}(\mathbf{u}^{H}[s](\boldsymbol{\Omega}-\lambda_{max}(\boldsymbol{\Omega}))\mathbf{I})\nonumber\\
&\mathbf{u}_{0}^{H}[s]\geq\gamma\sigma^{2},(\ref{pro21i}).&\label{pro66b}
\end{align}\label{pro66}%
\end{subequations}
\end{small}%
Since $\|\mathbf{u}[s]\|_{2}=1$, problem (\ref{pro66}) is reformulated as
\begin{small}
\begin{subequations}
\begin{align}
\min_{\mathbf{u}[s]}&~\frac{1}{S}\sum_{s=1}^{S}2\mathrm{Re}(\mathbf{u}^{H}[s](\boldsymbol{\Omega}-\lambda_{max}(\boldsymbol{\Omega}))\mathbf{I})\mathbf{u}_{0}^{H}[s],\label{pro67a}\\
\mbox{s.t.}~
&\lambda_{max}(\boldsymbol{\Omega}+2\mathrm{Re}(\mathbf{u}^{H}[s](\boldsymbol{\Omega}-\lambda_{max}(\boldsymbol{\Omega}))\mathbf{I})\mathbf{u}_{0}^{H}[s]\geq\gamma\sigma^{2},\nonumber\\
&(\ref{pro21h}),&\label{pro67b}
\end{align}\label{pro67}%
\end{subequations}
\end{small}%
The constraint condition in (\ref{pro21h}) is equivalently transformed as $\|\mathbf{u}\|_{2}\leq 1$, $\|\mathbf{u}\|_{2}\geq 1$, thus, we have
\begin{small}
\begin{subequations}
\begin{align}
\min_{\mathbf{u}[s]}&~\frac{1}{S}\sum_{s=1}^{S}2\mathrm{Re}(\mathbf{u}^{H}[s](\boldsymbol{\Omega}-\lambda_{max}(\boldsymbol{\Omega}))\mathbf{I})\mathbf{u}_{0}^{H}[s],\label{pro68a}\\
\mbox{s.t.}~
&\lambda_{max}(\boldsymbol{\Omega}+2\mathrm{Re}(\mathbf{u}^{H}[s](\boldsymbol{\Omega}-\lambda_{max}(\boldsymbol{\Omega}))\mathbf{I})\mathbf{u}_{0}^{H}[s]\geq\gamma\sigma^{2},&\label{pro68c}\\
&\|\mathbf{u}\|_{2}\leq 1, \|\mathbf{u}\|_{2}\geq 1,&\label{pro68b}
\end{align}\label{pro68}%
\end{subequations}
\end{small}%
$\|\mathbf{u}\|_{2}\geq 1$ can be approximated as $\|\mathbf{u}_{0}\|_{2}+2\mathrm{Re}(\mathbf{u}_{0}^{H}(\mathbf{u}-\mathbf{u}_{0}))\geq 1$. Problem (\ref{pro68}) is rewritten as
\begin{small}
\begin{subequations}
\begin{align}
\min_{\mathbf{u}[s]}&~\frac{1}{S}\sum_{s=1}^{S}2\mathrm{Re}(\mathbf{u}^{H}[s](\boldsymbol{\Omega}-\lambda_{max}(\boldsymbol{\Omega}))\mathbf{I})\mathbf{u}_{0}^{H}[s],\label{pro69a}\\
\mbox{s.t.}~
&(\ref{pro68c}),&\label{pro69b}\\
&\|\mathbf{u}\|_{2}\leq 1,&\label{pro69c}\\
&\|\mathbf{u}_{0}\|_{2}+2\mathrm{Re}(\mathbf{u}_{0}^{H}(\mathbf{u}-\mathbf{u}_{0}))\geq 1,&\label{pro69d}
\end{align}\label{pro69}%
\end{subequations}
\end{small}%
{which is convex, and can be solved using CVX}. Finally, the proposed {SCA} algorithm is summarized in \textbf{Algorithm~\ref{algo1}}
\begin{algorithm}%
\caption{Proposed {SCA} Optimization Algorithm for Problem (\ref{pro21})} \label{algo1}
\hspace*{0.02in}{\bf Initialize:}
$\bar{\mathbf{p}}_{u}^{(0)}[s]$, $\mathbf{w}^{(0)}[s]$, $\mathbf{u}^{(0)}[s]$.\\
\hspace*{0.02in}{\bf Repeat:}~$t=t+1$.\\
Given $\bar{\mathbf{p}}_{u}^{(t+1)}[s]$, $\mathbf{u}^{(t)}[s]$, {find optimal} solution $\mathbf{w}^{*}[s]$ of problem (\ref{pro202}). Update $\mathbf{w}^{(t+1)}[s]=\mathbf{w}^{*}[s]$;\\
Given $\mathbf{w}^{(t+1)}[s]$, $\mathbf{u}^{(t)}[s]$, {find optimal} solution $\bar{\mathbf{p}}_{u}^{*}[s]$ of problem (\ref{pro63}). Update $\bar{\mathbf{p}}_{u}^{(t+1)}[s]=\bar{\mathbf{p}}_{u}^{*}[s]$;\\
Given $\mathbf{w}^{(t+1)}[s]$, $\bar{\mathbf{p}}_{u}^{(t)}[s]$, {find optimal} solution $\mathbf{u}^{*}[s]$ of problem (\ref{pro69}). Update $\mathbf{u}^{(t+1)}[s]=\mathbf{u}^{*}[s]$;\\
\hspace*{0.02in}{\bf Until:}~$|R_{sum}^{(t+1)}[s]-R_{sum}^{(t)}[s]|\leq \nu$.\\
\hspace*{0.02in}{\bf Output:}
$\mathbf{w}^{(t+1)}[s]$, $\bar{\mathbf{p}}_{u}^{(t+1)}[s]$,$\mathbf{u}^{(t+1)}[s]$.\\
\end{algorithm}
In \textbf{Algorithm~\ref{algo1}}, \textcolor{blue}{the computational complexity of (\ref{pro202}) is $\mathcal{O}((3N_{t})^{3.5})$, and the computational complexity of (\ref{pro63}) is $\mathcal{O}((5(2+N_{t}))^{3.5})$\cite{10529955}.} The computational complexity of
receive beamforming in (\ref{pro69}) is $\mathcal{O}(N_{r}^{3})$\cite{7547360}. Thus, the total computational complexity of \textbf{Algorithm~\ref{algo1}} is calculated by $\mathcal{O}(I_{1}(I_{2}(3N_{t})^{3.5}+I_{3}(5(2+N_{t}))^{3.5}+I_{4}N_{r}^{3}))$,
where $I_{1}$, $I_{2}$, $I_{3}$ and $I_{4}$ denote the number of iterations in the
step $2$ to $5$.

\section{Numerical Results}\label{V}
\begin{table}[!ht]
\centering
\caption{Notations.}
\label{notations}
\begin{tabular}{cp{5.0cm}}
\toprule[1pt]
\textbf{Symbol} & \textbf{Descriptions} \\
\midrule
$N_{t}$ & 16. \\
$N_{r}$ &  8. \\
$T$ & 30s. \\
$S$ & 50. \\
$\Delta_{s}$ & $0.6$~s. \\
$\sigma_{d}^{2}$ & $-80$~dBm. \\
$\sigma_{t}^{2}$ & $-100$~dBm. \\
$\beta_{ud}=$ & $15$~dB. \\
$\beta_{ut}$ & $5$~dB. \\
$\sigma_{ut}^{2}$ & $5$~dB. \\
$D$ & $30$~m. \\
$P$ & $30$~dBm. \\
$v_{max}$ & $50$~m/s. \\
$\kappa$ & $3.1$. \\
\bottomrule[1pt]
\end{tabular}\label{TA1}
\end{table}

\begin{figure}[htbp]
\centering
\begin{minipage}[t]{0.45\textwidth}
\centering
\includegraphics[scale=0.4]{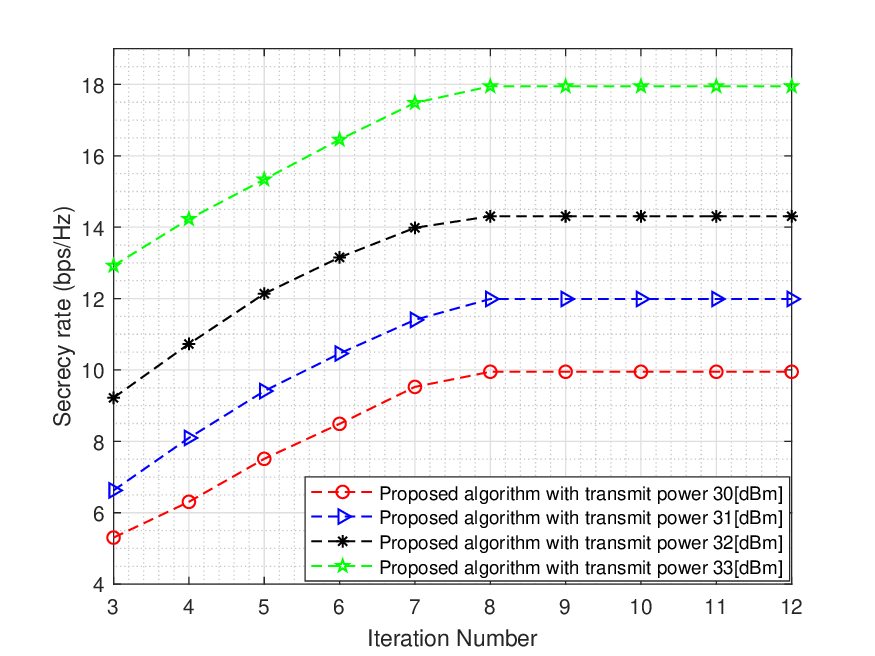}
\caption{{Average secrecy rate of ISAC-UAV system versus the number of iterations with different transmit powers.}}
\label{FIGURE1}
\end{minipage}
\end{figure}
{In this section, we present numerical examples to highlight the secrecy performance } of the ISAC-UAV communication system. 
The positions of IoT device and UT are set as $(10,20,0)$, $(30,30,0)$ in meters. The initial {UAV position} is $(0,0,15)$, and the final {UAV position} is $(60,30,15)$ in meters. The maximum speed of {the} UAV is $50~m/s$\cite{10529955}, slot length $\Delta_{s}=0.6$ s. The path loss coefficient and the
path loss exponents $\kappa=3.1$, $\alpha=1.5$ The Rician factor of UAV-to-IoT device $\beta_{ud}=15$ ~dB\cite{li2024uav}. The Rician factor of UAV-to-Untrusted target,UAV-to-IoT device, $\beta_{ut}=5$~ dB,  $\beta_{rt}=5$~dB. The max transmit power $P=30$~dBm. \textcolor{blue}{The simulation parameters are summarized in \textbf{Table}~\ref{TA1}.}

Fig.~\ref{FIGURE1} shows the convergence behavior of the proposed algorithm. 
The results show that all the curves increase monotonously and converge to a certain upper bound from the initial point. It can be seen from the figure that our proposed algorithm converges quickly after 8 iterations.
In addition, Fig.~\ref{FIGURE1} also shows that the variation of the average secrecy rate with transmit power. 
We can see that as the transmit power increases from $30$~dBm to $33$~dBm, the average security rate of the system {increases significantly from 10 to 18 bps/Hz}.

From Fig.~\ref{FIGURE6}, the UAV will first approach the IoT device, then remain at that location to maximise the average secrecy rate, before proceeding to the endpoint while satisfying the sensing rate for the untrusted target. Comparing Fig.~\ref{FIGURE6}(a) and Fig.~\ref{FIGURE6}(b), {we see} that when the sensing rate is increased from $\gamma=5$~bps/Hz to $10$~bps/Hz, the UAV must fly closer to the UT to satisfy the higher sensing rate constraint. However, this will result in a low secrecy rate for the IoT device. Comparing Fig.~\ref{FIGURE6}(b) and Fig.~\ref{FIGURE6}(c), when $\gamma$ is further increased from $10$~bps/Hz to $15$~bps/Hz, the UAV must fly even closer to the UT to satisfy the higher sensing rate constraint, which will results in a lower secrecy rate for the IoT device. This highlights a fundamental tradeoff in the secure ISAC-UAV system where the UAV-ISAC needs to balance the requirements of the sensing rate and secrecy rate based on the locations of the IoT device and untrusted target.
\begin{figure}[htbp]
\begin{minipage}[t]{0.45\textwidth}
\centering
\includegraphics[scale=0.4]{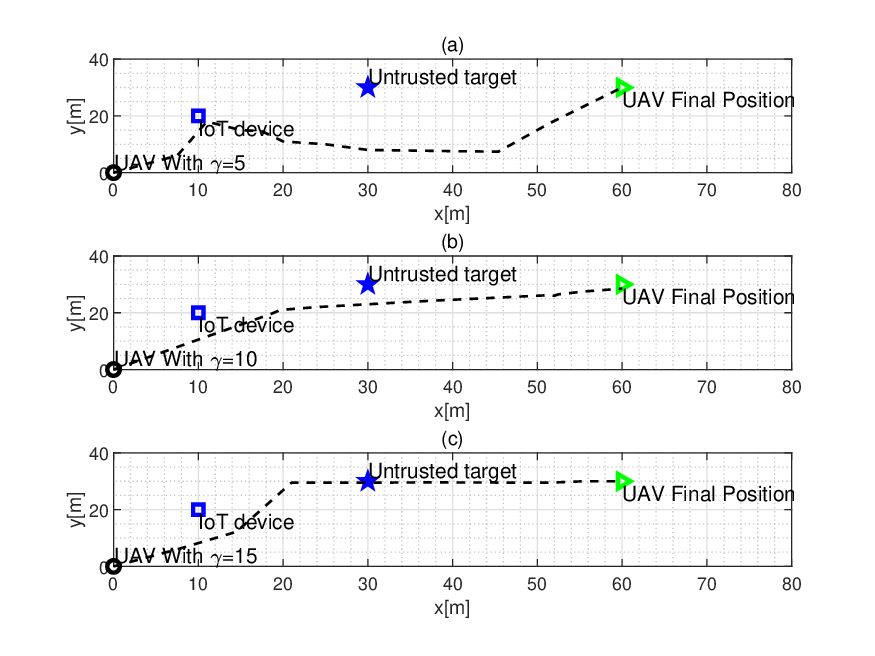}
\caption{\textcolor{blue}{UAV trajectory {optimization with different sensing rate constraints where $\gamma=5$~bps/Hz, and average secrecy rate is $15.7$~bps/Hz in (a), $\gamma=10$~bps/Hz, and average secrecy rate is $11.3$~bps/Hz in (b) and $\gamma=15$~bps/Hz, and average secrecy rate is $6.6$~bps/Hz in (c)}.}}
\label{FIGURE6}
\end{minipage}
\end{figure}\vspace{-10pt}

Fig.~\ref{FIGURE7} illustrates the secrecy rate of the ISAC-UAV system versus the number of time slots $S$ compared to the baseline schemes. Notably, the proposed algorithm achieves the highest secrecy rate compared to several baseline schemes. From Fig.~\ref{FIGURE7}, optimizing the UAV trajectory significantly benefits the transmit beamforming vector optimization in the ISAC-UAV system, especially over a sufficient number of mission cycles, such as $S\geq 10$. This improvement is primarily because the performance gains of the proposed algorithm mainly arise from the UAV trajectory design, which becomes more evident with a larger $S$. Moreover, the secrecy rate of the maximum ratio transmission (MRT) transmit beamforming optimization scheme is poor. \textcolor{blue}{This is because MRT  technology optimizes the antenna weights at the receiver to enhance signal transmission in a specific direction. If this improved signal is intercepted by an eavesdropper positioned in that direction, the higher gain results in the eavesdropper receiving a stronger signal.} Finally, based on Fig.~\ref{FIGURE7}, the proposed algorithm achieves a higher secrecy rate than the SDR-based and SCA-based algorithms in \cite{li2024uav} and \cite{10054167}. This is because the schemes in \cite{li2024uav} and \cite{10054167} deal with target sensing accuracy constraints based on CRB, and directly applying them in the sensing rate scenarios result in a worse secrecy rate.
\begin{figure}[htbp]
\begin{minipage}[t]{0.45\textwidth}
\centering
\includegraphics[scale=0.40]{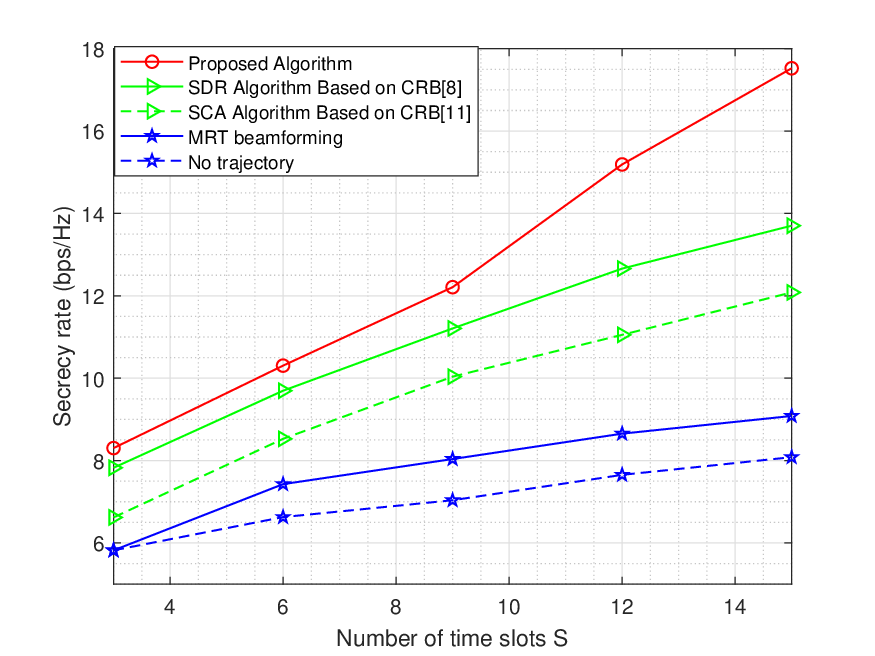}
\caption{Secrecy rate versus the time slot $S$.}
\label{FIGURE7}
\end{minipage}
\end{figure}\vspace{-10pt}

\begin{figure}[htbp]
\begin{minipage}[t]{0.45\textwidth}
\centering
\includegraphics[scale=0.40]{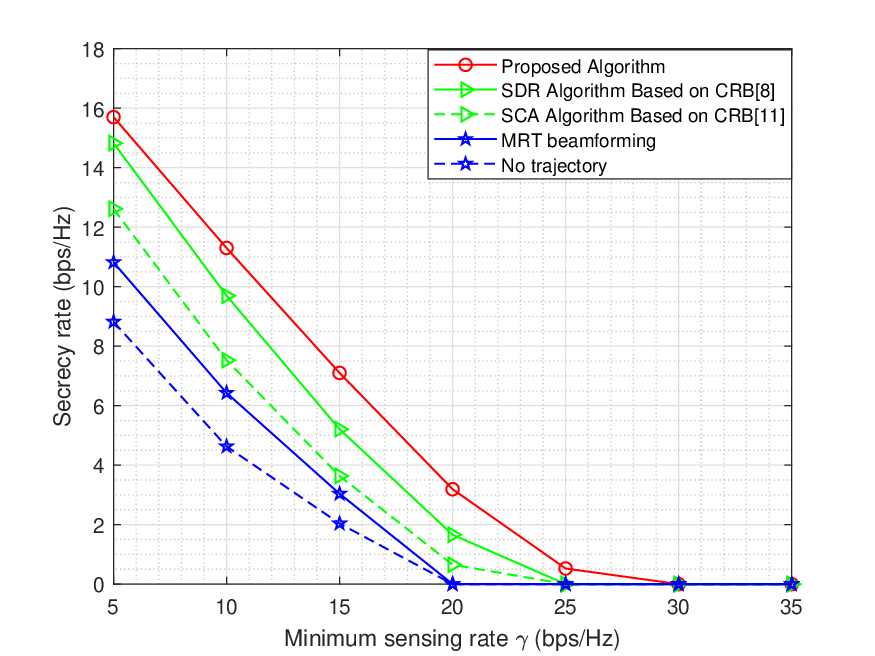}
\caption{Secrecy rate versus the minimum SNR $\gamma$.}
\label{FIGURE8}
\end{minipage}
\end{figure}\vspace{-10pt}

\textcolor{blue}{As shown in Fig.~\ref{FIGURE8}, the system's secure rate gradually decreases as the minimum sensing rate increases. With the higher sensing rate requirements, the system must allocate more power to satisfy the sensing rate constraint, thereby reducing the secure rate. However, we observe that the proposed algorithm exhibits the most minor decrease in security rate among all the algorithms. Additionally, as the sensing rate continues to increase, the system power eventually becomes insufficient to meet the minimum sensing rate constraint, at which point the system ceases to function correctly, and we set the secure rate to zero in our simulations.}

\section{Conclusion and Future Works}\label{VI}
In this paper, we {analysed the secrecy performance of a ISAC-UAV system communicating with a ground IoT device and sensing an untrusted target which is considered to be a potential eavesdropper.} 
{For this system,} we jointly optimize UAV trajectory and beamforming vectors to improve the secrecy rate while meeting the sensing rate constraint. Although the average secrecy rate optimization problem is a non-convex problem with multi-variable coupling, we propose {an efficient} algorithm to optimize the UAV trajectory, transmit beamforming, and receive beamforming to maximize the average secrecy rate of the ISAC-UAV system. Finally, we provide numerical results to evaluate the ISAC-UAV secrecy rate performance, showing important trade-offs between secure communication and sensing. The results show that our proposed method can balance the secrecy rate and target sensing constraints by optimizing UAV trajectories and beamforming vectors. \textcolor{blue}{Finally, this paper only considers a scenario with perfect CSI. However, it is challenging to obtain accurate CSI due to the mobility of UAVs. Therefore, in future work, we will design a robust transmission scheme for UAV-ISAC systems by integrating the minimum mean square error (MMSE) method with the proposed algorithm.}

\end{document}